\documentclass[ALICE,manyauthors]{cernphprep}
\usepackage[comma,square,numbers,sort&compress]{natbib}

\usepackage{lineno}
\usepackage{xspace}
\usepackage{hyperref}

\usepackage{multirow}
\usepackage{graphicx}
\usepackage{subfigure}
\usepackage{color}
\usepackage{dcolumn}
\usepackage{bm}
\usepackage[T1]{fontenc}
\usepackage{orcidlink}

\begin{document}
%

\newcommand{\comm}[1]{\textcolor{green}{#1}}

\newcommand{\pp}           {pp\xspace}
\newcommand{\ppbar}        {\mbox{$\mathrm {p\overline{p}}$}\xspace}
\newcommand{\XeXe}         {\mbox{Xe--Xe}\xspace}
\newcommand{\PbPb}         {\mbox{Pb--Pb}\xspace}
\newcommand{\pA}           {\mbox{pA}\xspace}
\newcommand{\pPb}          {\mbox{p--Pb}\xspace}
\newcommand{\AuAu}         {\mbox{Au--Au}\xspace}
\newcommand{\dAu}          {\mbox{d--Au}\xspace}

\newcommand{\s}            {\ensuremath{\sqrt{s}}\xspace}
\newcommand{\snn}          {\ensuremath{\sqrt{s_{\mathrm{NN}}}}\xspace}
\newcommand{\pt}           {\ensuremath{p_{\rm T}}\xspace}
\newcommand{\meanpt}       {$\langle p_{\mathrm{T}}\rangle$\xspace}
\newcommand{\ycms}         {\ensuremath{y_{\rm CMS}}\xspace}
\newcommand{\ylab}         {\ensuremath{y_{\rm lab}}\xspace}
\newcommand{\etarange}[1]  {\mbox{$\left | \eta \right |~<~#1$}}
\newcommand{\yrange}[1]    {\mbox{$\left | y \right |~<~#1$}}
\newcommand{\dndy}         {\ensuremath{\mathrm{d}N_\mathrm{ch}/\mathrm{d}y}\xspace}
\newcommand{\dndeta}       {\ensuremath{\mathrm{d}N_\mathrm{ch}/\mathrm{d}\eta}\xspace}
\newcommand{\avdndeta}     {\ensuremath{\langle\dndeta\rangle}\xspace}
\newcommand{\dNdy}         {\ensuremath{\mathrm{d}N_\mathrm{ch}/\mathrm{d}y}\xspace}
\newcommand{\Npart}        {\ensuremath{N_\mathrm{part}}\xspace}
\newcommand{\Ncoll}        {\ensuremath{N_\mathrm{coll}}\xspace}
\newcommand{\dEdx}         {\ensuremath{\textrm{d}E/\textrm{d}x}\xspace}
\newcommand{\RpPb}         {\ensuremath{R_{\rm pPb}}\xspace}
\newcommand{\dVdy}         {\ensuremath{\mathrm{d}V/\mathrm{d}y}\xspace}

\newcommand{\nineH}        {$\sqrt{s}~=~0.9$~Te\kern-.1emV\xspace}
\newcommand{\seven}        {$\sqrt{s}~=~7$~Te\kern-.1emV\xspace}
\newcommand{\twoH}         {$\sqrt{s}~=~0.2$~Te\kern-.1emV\xspace}
\newcommand{\twosevensix}  {$\sqrt{s}~=~2.76$~Te\kern-.1emV\xspace}
\newcommand{\five}         {$\sqrt{s}~=~5.02$~Te\kern-.1emV\xspace}
\newcommand{\twosevensixnn}{$\sqrt{s_{\mathrm{NN}}}~=~2.76$~Te\kern-.1emV\xspace}
\newcommand{\fivenn}       {$\sqrt{s_{\mathrm{NN}}}~=~5.02$~Te\kern-.1emV\xspace}
\newcommand{\LT}           {L{\'e}vy-Tsallis\xspace}
\newcommand{\GeVc}         {Ge\kern-.1emV/$c$\xspace}
\newcommand{\MeVc}         {Me\kern-.1emV/$c$\xspace}
\newcommand{\TeV}          {Te\kern-.1emV\xspace}
\newcommand{\GeV}          {Ge\kern-.1emV\xspace}
\newcommand{\MeV}          {Me\kern-.1emV\xspace}
\newcommand{\GeVmass}      {Ge\kern-.2emV/$c^2$\xspace}
\newcommand{\MeVmass}      {Me\kern-.2emV/$c^2$\xspace}
\newcommand{\lumi}         {\ensuremath{\mathcal{L}}\xspace}

\newcommand{\ITS}          {\rm{ITS}\xspace}
\newcommand{\TOF}          {\rm{TOF}\xspace}
\newcommand{\ZDC}          {\rm{ZDC}\xspace}
\newcommand{\ZDCs}         {\rm{ZDCs}\xspace}
\newcommand{\ZNA}          {\rm{ZNA}\xspace}
\newcommand{\ZNC}          {\rm{ZNC}\xspace}
\newcommand{\SPD}          {\rm{SPD}\xspace}
\newcommand{\SDD}          {\rm{SDD}\xspace}
\newcommand{\SSD}          {\rm{SSD}\xspace}
\newcommand{\TPC}          {\rm{TPC}\xspace}
\newcommand{\TRD}          {\rm{TRD}\xspace}
\newcommand{\VZERO}        {\rm{V0}\xspace}
\newcommand{\VZEROA}       {\rm{V0A}\xspace}
\newcommand{\VZEROC}       {\rm{V0C}\xspace}
\newcommand{\Vdecay} 	   {\ensuremath{V^{0}}\xspace}

\newcommand{\ee}           {\ensuremath{e^{+}e^{-}}} 
\newcommand{\pip}          {\ensuremath{\pi^{+}}\xspace}
\newcommand{\pim}          {\ensuremath{\pi^{-}}\xspace}
\newcommand{\kap}          {\ensuremath{\rm{K}^{+}}\xspace}
\newcommand{\kam}          {\ensuremath{\rm{K}^{-}}\xspace}
\newcommand{\pbar}         {\ensuremath{\rm\overline{p}}\xspace}
\newcommand{\kzero}        {\ensuremath{{\rm K}^{0}_{\rm{S}}}\xspace}
\newcommand{\lmb}          {\ensuremath{\Lambda}\xspace}
\newcommand{\almb}         {\ensuremath{\overline{\Lambda}}\xspace}
\newcommand{\Om}           {\ensuremath{\Omega^-}\xspace}
\newcommand{\Mo}           {\ensuremath{\overline{\Omega}^+}\xspace}
\newcommand{\X}            {\ensuremath{\Xi^-}\xspace}
\newcommand{\Ix}           {\ensuremath{\overline{\Xi}^+}\xspace}
\newcommand{\Xis}          {\ensuremath{\Xi^{\pm}}\xspace}
\newcommand{\Oms}          {\ensuremath{\Omega^{\pm}}\xspace}
\newcommand{\degree}       {\ensuremath{^{\rm o}}\xspace}
\newcommand{\Hee}          {\ensuremath{^{3}\mathrm{He}}\xspace}
\newcommand{\antiHee}
{\ensuremath{^{3}\overline{\mathrm{He}}}\xspace}
\newcommand{\HH}          {\ensuremath{\mathrm{t}}\xspace}
\newcommand{\antiHH}      {\ensuremath{\overline{\mathrm{t}}}\xspace}
\newcommand{\hyp}          {\ensuremath{^{3}_{\Lambda}\mathrm H}\xspace}
\newcommand{\antihyp}      {\ensuremath{^{3}_{\overline{\Lambda}} \overline{\mathrm H}}\xspace}
\newcommand{\he}          {\ensuremath{^{4}\mathrm{He}}\xspace}
\newcommand{\antihe}
{\ensuremath{^{4}\overline{\mathrm{He}}}\xspace}

\newcommand{\permille}{\textperthousand\xspace}
\newcommand{\mt}   {\ensuremath{m_{\mathrm{T}}}\xspace}
\newcommand{\dbar}   {\ensuremath{\overline{\mathrm{d}}}\xspace}
\newcommand{\dd}     {\ensuremath{\mathrm{d}}\xspace}
\newcommand{\gmom}   {\mbox{\ensuremath{\mathrm{GeV}/c}}\xspace}
\newcommand{\dedx}{\mbox{\ensuremath{\dd E/\dd x}}\xspace}
\newcommand{\dvdy}{\mbox{\ensuremath{\dd V/\dd y}}\xspace}

\begin{titlepage}
\PHyear{2023}       
\PHnumber{255}      
\PHdate{13 November}  

\title{Measurement of (anti)alpha production in central Pb--Pb collisions\\ at $\mathbf{\sqrt{\boldsymbol{s}_{\rm NN}}= 5.02}$ TeV}
\ShortTitle{(Anti)alpha production in Pb--Pb collisions}   

\Collaboration{ALICE Collaboration\thanks{See Appendix~\ref{app:collab} for the list of collaboration members}}
\ShortAuthor{ALICE Collaboration} 

\begin{abstract}

In this letter, measurements of (anti)alpha production in central (0--10\%) Pb--Pb collisions at a center-of-mass energy per nucleon--nucleon pair of \snn=~5.02 TeV are presented, including the first measurement of an antialpha transverse-momentum spectrum. Owing to its large mass, the production of (anti)alpha is expected to be sensitive to different particle production models. The production yields and transverse-momentum spectra of nuclei are of particular interest because they provide a stringent test of these models.
The averaged antialpha and alpha spectrum is compared to the spectra of lighter particles, by including it into a common blast-wave fit capturing the hydrodynamic-like flow of all particles. This fit is indicating that the (anti)alpha also participates in the collective expansion of the medium created in the collision. A blast-wave fit including only protons, (anti)alpha, and other light nuclei results in a similar flow velocity as the fit that includes all particles. 
A similar flow velocity, but a significantly larger kinetic freeze-out temperature is obtained when only protons and 
light nuclei are included in the fit. 
The coalescence parameter $B_4$ is well described by calculations from a statistical hadronization model but significantly underestimated by calculations assuming nucleus formation via coalescence of nucleons. 
Similarly, the (anti)alpha-to-proton ratio is well described by the statistical hadronization model. On the other hand, coalescence calculations including                 
approaches with different implementations of the (anti)alpha substructure tend to underestimate the data.
\end{abstract}
\end{titlepage}

\setcounter{page}{2} 


\section{Introduction} 

During the past five decades the production of light nuclei in heavy-ion reactions has been measured over a broad range of collision energies~\cite{PhysRevLett.37.667,Gosset:1976cy,PhysRevC.24.971,Nagamiya:1984vk,PhysRevC.28.1552,E886,PhysRevC.50.1024,PhysRevC.50.1077,PhysRevLett.74.2646,Bearden:1999iq,E864,Afanasiev200022,E896,PhysRevC.69.024902,Reisdorf:2010aa,PhysRevC.94.044906,HADES:2020lob,STAR:2020hya,STAR:2021ozh,STAR:2022hbp,STAR:2023uxk}. At center-of-mass energies of up to a few GeV light nucleus production is commonly understood in terms of nuclear break-up where the incoming nuclei disintegrate into lighter nuclear fragments. In contrast to this, the study of the production of antinuclei in heavy-ion collisions is a nascent field that emerged with the availability of heavy-ion colliders~\cite{SIMONGILLO1995483,Bearden:1999hv,E864-dbar,PhysRevLett.69.2345,PhysRevC.85.044913,Appelquist:1996qy,NA52:2002pxu,Ambrosini:2000qk,Arsenescu:2003mf,1367-2630-5-1-150,Adler:2001uy,Afanasiev:2007tv,PHENIX2005,Nygaard:2007zz,Abelev:2009ae,Arsene:2010px,Adam:2019wnb}. In particular, the antialpha was first observed only 13 years ago by the STAR Collaboration in Au--Au collisions at the Relativistic Heavy-Ion Collider (RHIC)~\cite{Agakishiev:2011ib}. At the Large Hadron Collider (LHC),
which provides the highest center-of-mass energies for heavy-ion collisions to date, measurements of the production of nuclei and antinuclei have so far mainly been performed by the
ALICE Collaboration in different collision systems~\cite{ALICE:2015rey,Adam:2015vda,Acharya:2017bso,ALICE:2017xrp,ALICE:2019dgz,ALICE:2019bnp,ALICE:2020foi,ALICE:2020zhb,ALICE:2020hjy,ALICE:2021mfm,ALICE:2021ovi,ALICE:2022zuz,ALICE:2022amd,ALICE:2022wpn,Nuclei_5TeV,ALICE:2022jmr,ALICE:2022weh,ALICE:2023aod}. Understanding the production mechanism of nuclei and antinuclei in ultrarelativistic collisions could provide deeper insights into the hadronization process and the quantum properties of composite hadronic systems.

Two different approaches exist that describe the production of light (anti)nuclei in heavy-ion collisions.
In statistical hadronization models (SHMs), often simply called thermal models, the production of hadrons and nuclei is described in the
framework of a grand-canonical ensemble employing only three parameters: temperature $T$, 
volume $V$, and baryo-chemical potential $\mu_B$~\cite{BraunMunzinger:1994iq,BraunMunzinger:2003zd,Andronic:2010qu,Braun-Munzinger:2018hat,Andronic:2017pug,Vovchenko:2018fiy,Donigus:2020ohq}. Previous measurements of the production of light (anti)nuclei in central Pb--Pb collisions~\footnote{Centrality in heavy-ion collisions is normally given in the inverse percentage of the overlap between the area of the collided nuclei, i.e. 0--10\% central collisions corresponds to the events where the collisions are mostly head-on and 80--90\% would be a peripheral collision where the colliding nuclei only have a small overlap. Semicentral corresponds to a centrality of 30--50\%.} by the ALICE Collaboration agreed well with a common SHM
fit to all available hadron and nucleus measurements with a temperature of T = (156.5$\pm$1.5)\,MeV and a baryo-chemical
potential of $\mu_B$ = (0.7$\pm$3.8)\,MeV~\cite{Andronic:2017pug}. The temperature is commonly understood in terms of a chemical freeze-out temperature
$T_{\mathrm{ch}}$ at which the abundances of hadrons and nuclei are fixed during the evolution of the fireball created in central Pb--Pb collisions~\cite{Braun-Munzinger:2018hat,ALICE:2022wpn}. It is compatible 
with the (pseudo)critical temperature $T_c$ predicted by the lattice QCD calculations for the transition between a hadronic
system and a quark--gluon plasma (QGP) at vanishing $\mu_B$~\cite{HotQCD:2018pds,Borsanyi:2020fev}. The interpretation in the context of the production of nuclei,
however, is not straightforward because significant modifications of the abundances of nuclei are expected from density and cross section arguments due to inelastic processes in the subsequent fireball evolution, often called hadronic phase, below $T_{\mathrm{ch}}$~\cite{Braun-Munzinger:2018hat,Vovchenko:2019aoz,Vovchenko:2022xil}.
In another class of models, nucleus formation is conjectured via the coalescence of nucleons in the final state 
of the system evolution~\cite{Butler:1963pp,Kapusta:1980zz,Csernai:1986qf,Steinheimer:2012tb}. The coalescence process is typically associated with the kinetic freeze-out temperature $T_{\rm kin}$, which corresponds to the temperature where the inelastic collisions cease and the (transverse) momentum spectra of the particles are frozen~\cite{Braun-Munzinger:2018hat,Sun:2022xjr}. The invariant yield $E_A\frac{\dd^3N_A}{\dd p^3_A}$ of nuclei with mass number A is connected to the 
final-state momentum distribution of protons $E_{\mathrm{p}}\frac{\dd^3N_{\mathrm{p}}}{\dd p^3_{\mathrm{p}}}$
via the coalescence parameter~\cite{Csernai:1986qf}:

\begin{equation}
	B_A=E_A\frac{\dd^3N_A}{\dd p^3_A}\left(E_{\mathrm{p}}\frac{\dd^3N_{\mathrm{p}}}{\dd p^3_{\mathrm{p}}}\right)^{-A},
	\label{eq:coalescence}
\end{equation}

assuming that protons and neutrons are produced in equal amounts at ultrarelativistic collision energies since both belong to the same isospin doublet.
The coalescence prescription can thus be employed to deduce the formation of nuclei based on measured proton yields 
as well as on nucleon distributions from event generators such as PYTHIA~\cite{Sjostrand:2014zea,Bierlich:2022pfr} and EPOS~\cite{Pierog:2013ria}, or transport models like UrQMD~\cite{Bass:1998ca,Sombun:2018yqh,Reichert:2022mek} or SMASH~\cite{Oliinychenko:2018odl,Oliinychenko:2018ugs,Petersen:2018jag}.

In pertinent formulations of the coalescence model, the coalescence probability incorporates a dependence 
on the spatial distribution of the nucleons at kinetic freeze-out and its overlap with the internal wave function of the 
nuclear cluster, leading to a characteristic dependence of $B_A$ and consequently the production yield of nuclei on the size of the collision system~\cite{Mrowczynski:1987oid,Mrowczynski:1989jd,Mrowczynski:1992gc,Scheibl:1998tk,Blum:2017iwq,Sun:2017ooe,Sun:2018mqq,Blum:2019suo,Bellini:2018epz,Bellini:2019zqc,Bellini:2020cbj}. 
This motivated detailed studies of
nuclear formation in pp, p--Pb, and non-central Pb--Pb collisions, where the yield ratio of nucleus $A$ relative to protons 
is studied as a function of the average charged-particle multiplicity per unit of pseudorapidity, $\langle {\rm d}N_{\rm ch}/{\rm d}\eta \rangle$. 
Indeed, the present data tend to confirm the system-size dependence predicted by coalescence models for the yield ratios deuteron to proton (d/p), triton to proton (t/p), and $^{3}$He to proton ($^{3}$He/p)
in small collision systems~\cite{Adam:2015vda,ALICE:2017xrp,ALICE:2019bnp,ALICE:2019dgz,ALICE:2020foi,ALICE:2020hjy,ALICE:2021mfm,ALICE:2021ovi}.
On the other hand, the statistical description of particle production in small collision systems requires a canonical formulation of the
statistical hadronization model, leading to the Canonical Statistical Model (CSM). 
This formulation entails an additional model parameter, the so-called correlation volume $V_{\rm C}$, inside which electric charge $Q$, strangeness $S$, and baryon number $B$ are conserved exactly~\cite{Becattini:2000jw,Sharma:2018jqf,Vovchenko:2018fiy,Vovchenko:2019kes,Donigus:2020ohq,Sharma:2022poi}. CSM calculations of nucleus-to-proton ratios result
in a suppression of the production of nuclei in small systems that is qualitatively compatible with the patterns observed in data,
but still tends to overestimate the yields of nuclei for realistic assumptions of $V_{\rm C}$~\cite{Vovchenko:2018fiy,Vovchenko:2019kes,ALICE:2022amd}.

In central and semi-central Pb--Pb collisions, recent results for d/p and $^{3}$He/p are compatible with both statistical hadronization
and coalescence models, while t/p in Pb--Pb is significantly closer to the coalescence model~\cite{Nuclei_5TeV}. It should be noted, however, 
that the yield of nuclei in Pb--Pb collisions may also be modified by absorption effects during the hadronic phase, as indicated by calculations from the UrQMD hybrid coalescence model~\cite{Reichert:2022mek}. 

The observed stiffening of transverse-momentum (\pt) spectra of hadrons produced in heavy-ion collisions can be interpreted 
in terms of a common radial flow field, arising from hydrodynamic expansion. The so-called blast-wave model~\cite{Schnedermann:1993ws} describes the radial boost of the light-flavor hadrons and nuclei 
arising from hydrodynamic expansion 
with a common set of parameters: the kinetic freeze-out temperature $T_{\rm kin}$, the mean radial expansion velocity $\langle\beta\rangle$, and 
an exponent $n$ of the radial velocity profile.
The measured \pt spectra are fitted with the Boltzmann-Gibbs blast-wave function~\cite{Schnedermann:1993ws}:

   \begin{equation}
     E\frac{{\rm d}^3N}{{\rm d}p^3} \propto \int_{0}^{R} \mt I_0 \left(\frac{\pt \sinh(\rho(r))}{T_{\rm kin}}\right) K_1 \left(\frac{\mt\cosh(\rho(r))}{T_{\rm kin}}\right) r\,{\rm d}r
     \label{eq:bwfit}
   \end{equation}

where \mt is the transverse mass ($\mt = \sqrt{m^2+{\pt}^2}$), $I_{0}$ and $K_{1}$ are the modified Bessel functions, and $\rho$ is the velocity profile given by:

   \begin{equation}
     \rho(r) = \tanh^{-1} \beta(r)=\tanh^{-1} \bigg[ \bigg( \frac{r}{R} \bigg)^{n}\beta_{\rm {max}} \bigg],
   \end{equation}

 where $r$ is the radial distance in the transverse plane, $R$ is the radius of the fireball and $\beta_{\rm max}$ is the transverse expansion velocity at the surface of the expanding fireball.

The study of (anti)alpha production in central Pb--Pb collisions is particularly interesting because it is the heaviest 
nucleus measured at the LHC so far.
In the SHM, there is a strong mass dependence of the statistical penalty factor (yield suppression when a baryon is added to the system) for typical values of $T_{\rm ch}$ and $\mu_{\rm B}$,
while predictions of coalescence models depend on nucleon densities and geometrical factors to the power of $A$. For the SHM the antialpha-to-alpha ratio is expected to be proportional to $\exp(-8\mu_\textrm{B}/T_{\rm ch})$~\cite{Cleymans:2011pe}, which is the strongest dependence of the thermal model parameters on the baryon number $A$. The microscopic coalescence models directly have problems to get the predictions, since they need much more nucleons to be produced initially that can be used then in the coalescence process. This makes the (anti)alpha a very sensitive probe for stringent tests of the production models of light nuclei.
In addition to already discussed models, there is also the idea that correlations are present already in the vacuum, allowing an antinucleus like antialpha to be directly excited from the vacuum~\cite{Heinz:1985pm,Greiner:2001ht,Greiner:2003db,Greiner:2012zz}. This would mean that the antinulcei rate could be much larger than the values predicted by SHM or coalescence models, which expect them to be rather equal.
Previous measurements of the integrated yields of antialpha and alpha in central Pb--Pb collisions at a center-of-mass energy per nucleon--nucleon pair of
$\sqrt{s_{\rm NN}}$ = 2.76 TeV agreed with a global fit of the SHM to the yields of all measured hadrons and nuclei~\cite{Acharya:2017bso}. No predictions for $A$ = 4 from coalescence models existed at the time.

In this letter, we present results on (anti)alpha production in central Pb--Pb collisions at \mbox{$\sqrt{s_{\rm NN}}$ = 5.02 TeV},
including the first ever measured transverse-momentum spectrum of the antialpha. The results are compared to
predictions by coalescence and statistical hadronization models. Together with previous results for different hadron
species and lighter nuclei, the \pt spectra are analyzed employing the blast-wave model. Throughout this letter, especially in the figures but also at some occasions in the text, \he instead of alpha is stated, which are used as equivalent. Note that with \he not the chemical element with electron shell but the \he nucleus is meant. 

In Sec.~\ref{sec:ana} the analysis is described, followed by the presentation of the systematic uncertainties in Sec.~\ref{sec:syst}. The results are discussed in Sec.~\ref{sec:res} and the conclusion is given in Sec.~\ref{sec:con}.

\section{Data analysis}
\label{sec:ana}

\subsection{Data sample and experimental apparatus}
The presented results are based on a data set of \PbPb collisions at $\mathbf{\sqrt{{\textit s}_{\rm NN}}} = 5.02$ TeV, collected in 2018, where $99.5\times10^6$ events in the 0--10$\%$ centrality interval~\cite{alice_centrality} were analyzed. These events are the sum of a minimum-bias trigger of lower bandwidth and a central trigger with a higher bandwidth, giving $12.6\times10^6$ and $86.9\times10^6$ events, respectively. The number of antialpha candidates outside the chosen interval is negligible and would not give any statistical benefit.

The ALICE apparatus~\cite{Aamodt:2008zz,Abelev:2014ffa} provides excellent particle identification and vertexing capabilities.
The \mbox{(anti)}alpha was reconstructed and identified using the Inner Tracking System (ITS), the Time Projection Chamber (TPC), the Transition Radiation Detector (TRD), and the Time-Of-Flight detector (TOF).
These detectors are all located inside a homogeneous magnetic field with a strength of 0.5~T and cover the full azimuth in the pseudorapidity range $|\eta|<0.9$. Interactions located inside $|z|<$ 10 cm are selected, where $z$ is the distance from the nominal interaction point along the beam direction.

The ITS~\cite{ITS2010} is a silicon detector consisting of six cylindrical layers. It is used for charged-particle tracking and for the reconstruction of primary and secondary vertices. It can also be used to separate primary nuclei from secondary, knocked-out nuclei from the detector material, via the distance of closest approach (DCA) of the track to the primary vertex.

The TPC~\cite{Alme2010} is the main tracking device of the ALICE apparatus. It is a gas-filled cylinder and provides charged-particle tracking and particle identification via the specific energy loss per path length (d$\textit E$/d$\textit x$) with a resolution of 6\% in Pb--Pb collisions. 

The TOF detector~\cite{TOF2013} provides identification of light (anti)nuclei by means of the velocity determination from the calculated path length of the track and the time-of-flight measurement. Its total time resolution for tracks in \PbPb collisions corresponds to about 65 ps which is determined by the intrinsic time resolution of the detector and the resolution of the event collision time measurement.
By combining TPC and TOF information, (anti)alphas can be identified from $\pt = 2~\gmom$ up to $6~\gmom$ in \PbPb collisions. 

The TRD~\cite{ALICE:2017ymw} can be used to improve the momentum resolution and significantly reduces the probability of random matches between tracks and TOF hits.

The V0 detectors~\cite{VZERO2013} measure the arrival time of particles with a resolution of 1 ns, by utilizing a pair of forward and backward scintillator arrays (covering the pseudorapidity ranges $2.8<\eta<5.1$ and $-3.7<\eta<-1.7$). They are used for triggering purposes and for rejection of beam--gas interactions. In addition, they provide the centrality trigger in Pb--Pb collisions~\cite{alice_centrality}, and they are also used for offline centrality determination. 

The Zero Degree Calorimeter (ZDC) consists of two sets of hadronic calorimeters, which are located 112.5\,m away from the interaction point on both sides of it, and of one set of electromagnetic calorimeters, placed 7 m away from the interaction point~\cite{Aamodt:2008zz} on one side of it. It is located at 0$^{\circ}$ relative to the beam direction. 

\subsection{Event and track selection}
\label{evtrkSel}

The data were collected using a minimum-bias trigger requiring at least one hit in both V0 detectors. In addition, a trigger on central collisions was used, also determined by the V0 detectors,  selecting collisions in the 0--10\% centrality interval.
To reject the events triggered by the interactions of the beam with the residual gas in the LHC vacuum pipe, the timing information of the V0 scintillator arrays is used. A further selection using the ZDC is applied in order to reject the electromagnetic beam--beam interactions and beam--satellite bunch collisions~\cite{Herr:941319}. This is done by selecting good events from the correlation between the sum and the difference of arrival times measured in each of the ZDCs~\cite{Abelev:2014ffa}. All these rejection steps are done in the offline analysis.

The production yield of (anti)alphas is measured at midrapidity ($|y|\,<$ 0.5). 
Only tracks in the full tracking acceptance of $|\eta|\,<$ 0.8 are selected. 
In order to guarantee good track momentum and d$\textit E$/d$\textit x$ resolution in the relevant $p_{\mathrm{T}}$ ranges, the selected tracks are required to have at least 70 out of 159 possible reconstructed points in the TPC and at least two points in the ITS out of which at least one is in the two innermost layers, the Silicon Pixel Detector (SPD). 
The requirement of at least one point in the SPD assures a resolution better than 300 $\mu$m on the distance of closest approach to the primary vertex 
for the selected tracks~\cite{Abelev:2014ffa}.
Furthermore, it is required that the $\chi^{2}$ per TPC reconstructed point is less than 2.5 and tracks with a kink, which originate from weak decays, where the decay products are one charged and at least one neutral particle, are rejected.

\subsection{Particle identification}
\label{partId}

\begin{figure}[!tbh]
    \begin{center}
    \includegraphics[width = 0.7\textwidth]{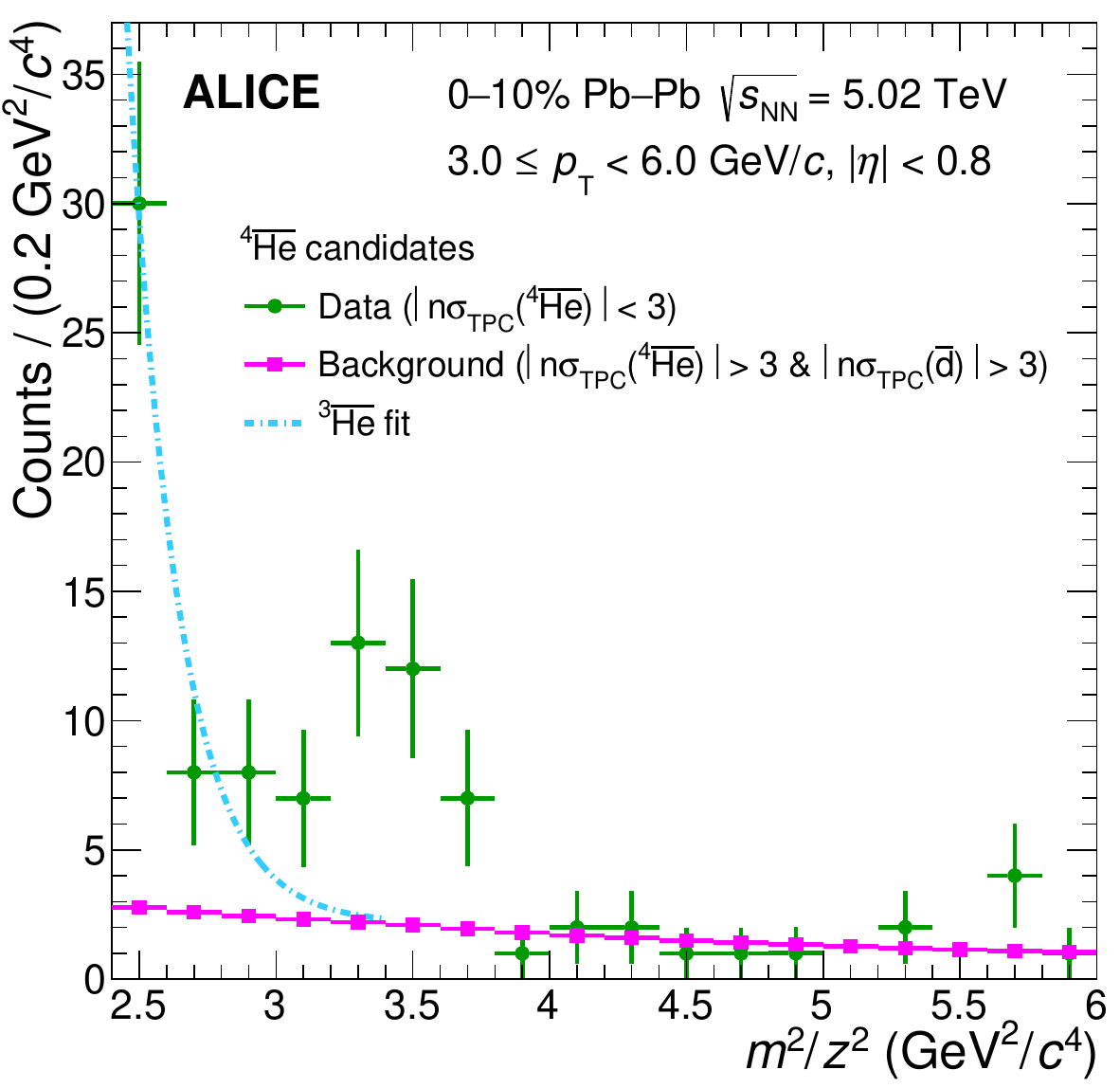}
    \end{center}
    \caption{(Color online) $m^2/z^2$ distribution for ${}^{4}\overline{{\rm He}}$ candidates (green). The background (magenta) is constructed by selecting all candidates outside the \antihe TPC window of 3\,$\sigma$ around the alpha mass hypothesis ($|n\sigma_{\rm TPC}({}^{4}\overline{{\rm He}})| > 3$) and in addition outside the 3\,$\sigma$ window around the deuteron mass hypothesis ($|n\sigma_{\rm TPC}(\overline{{\rm d}})| > 3$). The blue line is an exponential fit to the rise at lower masses originating from ${}^{3}\overline{{\rm He}}$ candidates.}
    \label{fig:mass}
\end{figure}

Particles with electric charge $z=2$ are well separated in the TPC from the particles with $z=1$, as they have a four times larger specific energy loss (d$E$/d$x$). However, to distinguish the alphas from the much more abundant $^3$He (by a factor of the order of 10$^3$) the d$E$/d$x$ information is combined with the mass calculated from the time-of-flight measured with the TOF and the track momentum. The energy loss in the TPC can be described by the Bethe--Bloch formula for a given mass hypothesis.
To select the (anti)alphas it is required that the energy loss of the track lies in a 3\,$\sigma$ window around the expected values for alpha particles, where $\sigma$ is the d$E$/d$x$ resolution. In addition, it is required that the track is matched to a hit in the TOF detector. Figure~\ref{fig:mass} shows the $m^2/z^2$ distribution of the TOF detector for antialpha candidates (green) in the \pt interval between 3 and 6\,GeV/$c$. The $m^2/z^2$ for true (anti)alphas is 3.475\,GeV$^2$/$c^4$. Note that in the $m^2/z^2$ distributions, $^{4}$He are clearly separated from $^{3}$He, for which $m^2/z^2$ is 2.0\,GeV$^2$/$c^4$.
The background (magenta) is coming from TOF mismatches, which is the case if a track in the TPC is associated with the wrong hit in the TOF detector, resulting in a wrong mass. To describe the background a data-driven approach with only one free parameter is used. The background is determined by selecting all tracks in the TPC outside a 3\,$\sigma$ interval of the expected Bethe--Bloch curve for alpha particles and in addition outside a 3\,$\sigma$ interval of the expected curve for the deuteron mass hypothesis, as alphas and deuterons have similar $m^2/z^2$. The background is then scaled to the height of the \he histogram by normalizing to the sideband on the right of the \he peak between 4.4 and 6\,GeV$^2/c^4$ and subtracted. This is done in each \pt interval separately except for the first \pt interval of the \antihe (2--3\,\gmom), where there is no background. The $^3$He contribution under the \he peak is described by an exponential fit to the tail of the $^3$He peak (blue dashed line). This is done in one \pt interval from 3 to 6\,\gmom and an (anti)$^3$He fraction (3\% for $^3\overline{\mathrm{He}}$ and 9\% for $^3$He) is determined for particles and antiparticles separately, which is then subtracted in each \pt interval individually. This is needed since the $^3$He contribution cannot be determined in each \pt interval separately due to the limited statistics. The (anti)alpha signal is counted in every \pt interval between 3 and 4.2\,GeV$^2/c^4$ due to the asymmetric shape of the signal in $m^2/z^2$. 

The \antihe raw yield is extracted in four \pt intervals between 2 and 6\,GeV/$c$. The $^4$He raw yield is only extracted in three \pt intervals between 3 and 6\,GeV/$c$, due to the large contribution of knocked-out alphas from the detector material and the support structure at low \pt. This contribution can only be extracted properly from data to Monte Carlo comparison  and is done usually in template fits in slices of \pt in the variable distance-of-closest approach. Unfortunately, this extraction is not possible for the presented analysis due to the small number of candidates. Nevertheless,
the comparison of the raw counts in \pt intervals suggests that the knock-out is negligible already for \pt $> 3$ GeV/$c$, since the raw yields of alpha and antialpha become similar in number.
For the statistical uncertainties of the data points the Poisson statistics is used.

\subsection{Corrections to the spectra}
\label{corrections}

The transverse-momentum spectra of the (anti)alphas are obtained by correcting the raw yields in the different \pt intervals of the analysis for tracking efficiency and detector acceptance. This is done by using Monte Carlo events, simulated with the HIJING event generator~\cite{Wang:1991hta}. As HIJING does not include (anti)alphas, they are injected into the event with flat distributions in \pt (between 0 and 10\,\gmom), azimuth (between 0 and 2$\pi$), and rapidity ($|y|\,<\,1$). 
The GEANT\,4~\cite{Agostinelli:2002hh} transport code is used to propagate the generated particles through a full simulation of the ALICE detector setup. The combined acceptance$\times$efficiency was determined for the (anti)alphas in the 0--10$\%$ centrality interval. As the nuclei are not produced with flat \pt distribution, the acceptance$\times$efficiency was weighted with a blast-wave shape applying an iterative method (see e.g.~Ref.~\cite{Nuclei_5TeV}), where the blast-wave parameters were taken from a fit to the (anti)alpha spectra.

\section{Systematic uncertainties}
\label{sec:syst}

To estimate the systematic uncertainties, different sources affecting the (anti)alpha measurement were studied, which are described in the following. Unless specified otherwise, all uncertainties are taken for all \pt intervals equally.

The first considered source of systematic uncertainty is related to possible imperfections in the description of the track reconstruction efficiency in the Monte Carlo simulations, which is usually estimated by varying the track selection criteria and by comparing the probability of attaching ITS hits to a TPC track (matching efficiency) in the data and in the simulation.
Owing to the low number of counts of the (anti)alpha analysis the systematic variations of the track selection criteria were found to be not significant within the statistical uncertainties by applying the check proposed by Barlow~\cite{Barlow:2002yb}. Therefore, the method based on varying the selections could not be used and instead systematic uncertainties based on similar studies of identified charged particles were assigned, namely 5\% for the TPC--ITS matching efficiency for all \pt intervals~\cite{ALICE:2019hno}. 

For the signal extraction, a systematic uncertainty between 6\% and 22\% for the \antihe and between 9\% and 14\% for the \he has been evaluated. This  uncertainty takes into account variations in fit functions and fit ranges used for yield extraction.

The limited knowledge of the interaction of (anti)nuclei with the detector material leads to another large contribution to the systematic uncertainties. The hadronic interaction cross section implemented in GEANT\,4~\cite{Agostinelli:2002hh,Grichine:2009zz,Uzhinsky:2011zz,Allison:2016lfl} is used to determine the acceptance$\times$efficiency. As there is no measurement of the \antihe inelastic interaction cross section so far, an uncertainty of 7\% is assumed, as done for the \antihe measured in the Pb--Pb data sample at $\snn = 2.76$\,TeV~\cite{Acharya:2017bso}. The 7\% are supposed to cover the difference between GEANT4, which was used for the propagation of the tracks in the detector material, and the true interaction cross sections~\cite{Acharya:2017bso}. This uncertainty represents the difference between the cross section implemented in GEANT4 to the one implemented in the AMS model in the rigidity interval where ALICE and AMS measurements overlap.

The material budget of the ALICE apparatus employed in the MC simulation was varied by $\pm$ 4.5\%, corresponding to the uncertainty of the ALICE material budget determination~\cite{Abelev:2014ffa}. This results in an uncertainty on the (anti)alpha spectra of 2$\%$.

The blast-wave weighting of the acceptance$\times$efficiency only affects the first \pt interval of the \antihe spectrum and the uncertainty was determined to be 3$\%$. This is half of the difference to the case when no blast-wave weighting is taken into account.

As there is a contribution of feed-down to the (anti)alphas from the decay of $^4_{\Lambda}$H and $^4_{\Lambda}\overline{\mathrm{H}}$, an additional uncertainty of 3$\%$ for particles and antiparticles in all \pt intervals was taken into account, estimated from a Monte Carlo study where these hypernuclei have been injected.

In total, all these contributions result in a systematic uncertainty between 12$\%$ and 24$\%$ for \antihe and between 12$\%$ and 16$\%$ for \he when added in quadrature. 

Most of the systematic uncertainties are correlated between \he and \antihe . The uncorrelated contributions are the uncertainty coming from the inelastic interaction cross section as well as the uncertainties on the background subtraction and the (anti)$^3$He contribution, which are part of the signal extraction.

\section{Results}
\label{sec:res}

\begin{figure}[!h]
    \begin{center}
    \includegraphics[width = 0.7\textwidth]{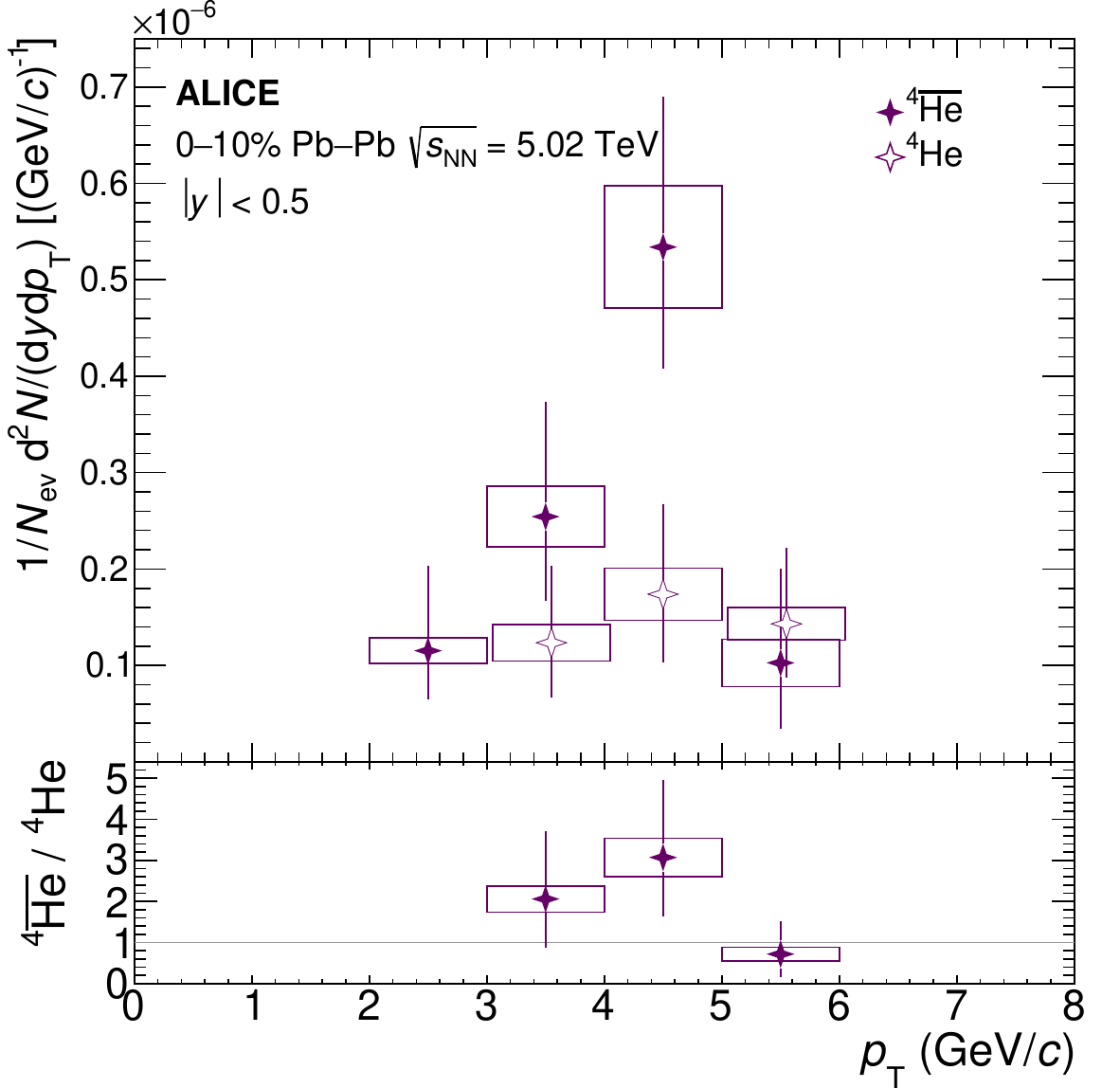}
    \end{center}
    \caption{Measured transverse-momentum distributions of ${}^{4}$He and ${}^{4}\overline{{\rm He}}$ (upper panel). The vertical lines indicate the statistical uncertainties, while the boxes represent the systematic ones. In the case that the statistical uncertainties would overlap the ${}^{4}$He points are a bit shifted on the x-axis.  The lower panel shows the ratio between ${}^{4}\overline{{\rm He}}$ and ${}^{4}$He with statistical and uncorrelated systematic uncertainties as the correlated systematic uncertainties cancel.}
    \label{fig:spectra}
\end{figure}

The size of the data sample presented in this letter exceeds that of a previous measurement in Pb--Pb collisions at \snn = 2.76 TeV~\cite{Acharya:2017bso} 
by about a factor of five. This allows for the determination of the transverse-momentum spectra for alpha and
antialpha, as shown in Fig.~\ref{fig:spectra}. In the case of the antialpha, this is the first ever measurement of the \pt distribution. In the \pt interval between 4 and 5\,\gmom there is a 2\,$\sigma$ discrepancy between particle and antiparticle yields, relative to the combination of statistical and systematic uncertainties, while in the other \pt intervals the alpha and antialpha yields are consistent within statistical uncertainties.
The antialpha-to-alpha ratio is shown in the lower panel of Fig.~\ref{fig:spectra}, where the error bars represent the statistical uncertainties and the boxes represent the uncorrelated systematic uncertainties, as the correlated ones cancel.
Both spectra were combined for further analysis by constructing the weighted average of the data points, where statistical and systematic uncertainties were considered.

\begin{figure}[!tbh]
    \begin{center}
    \includegraphics[width = 0.49\textwidth]{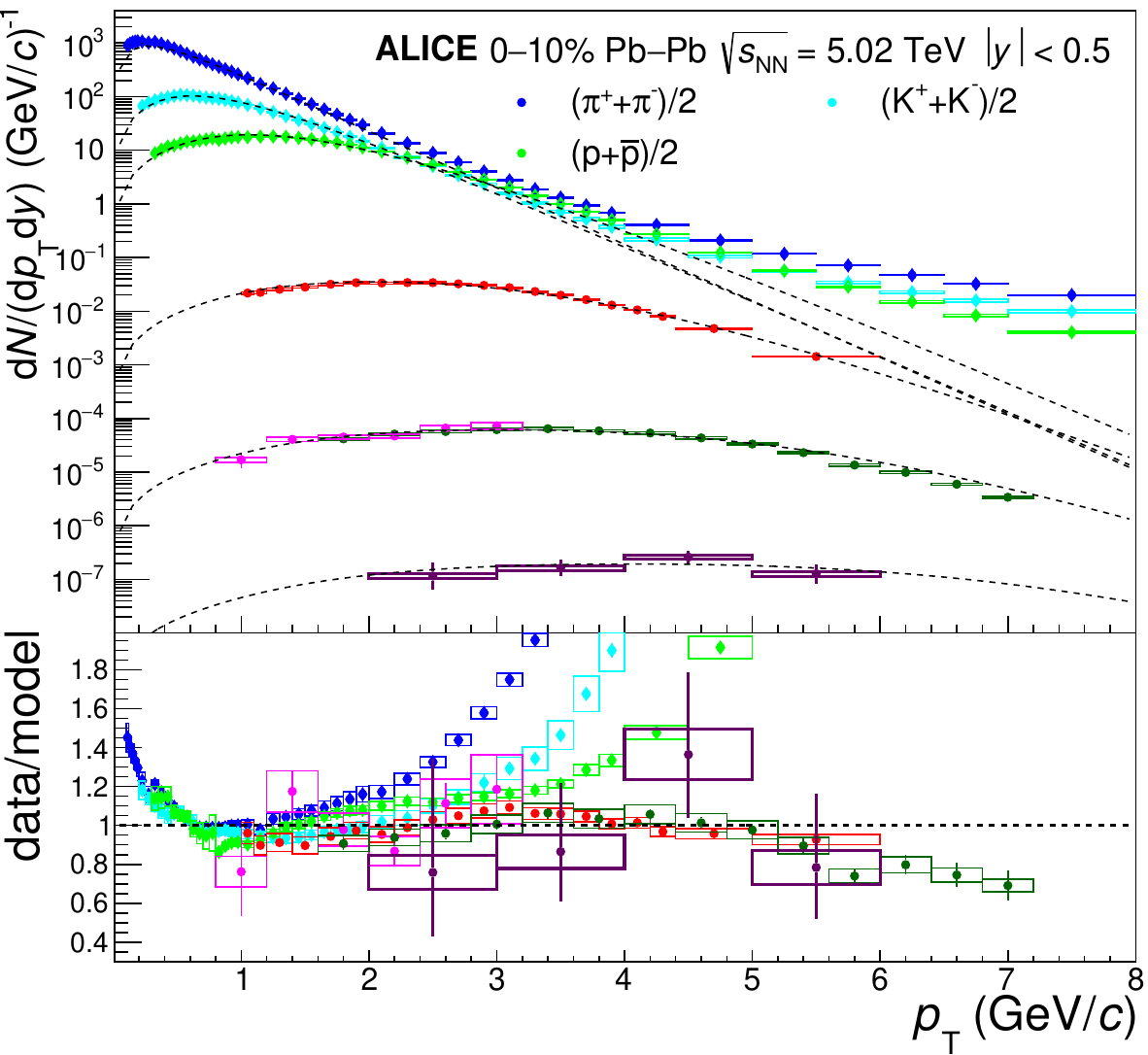}
    \includegraphics[width = 0.49\textwidth]{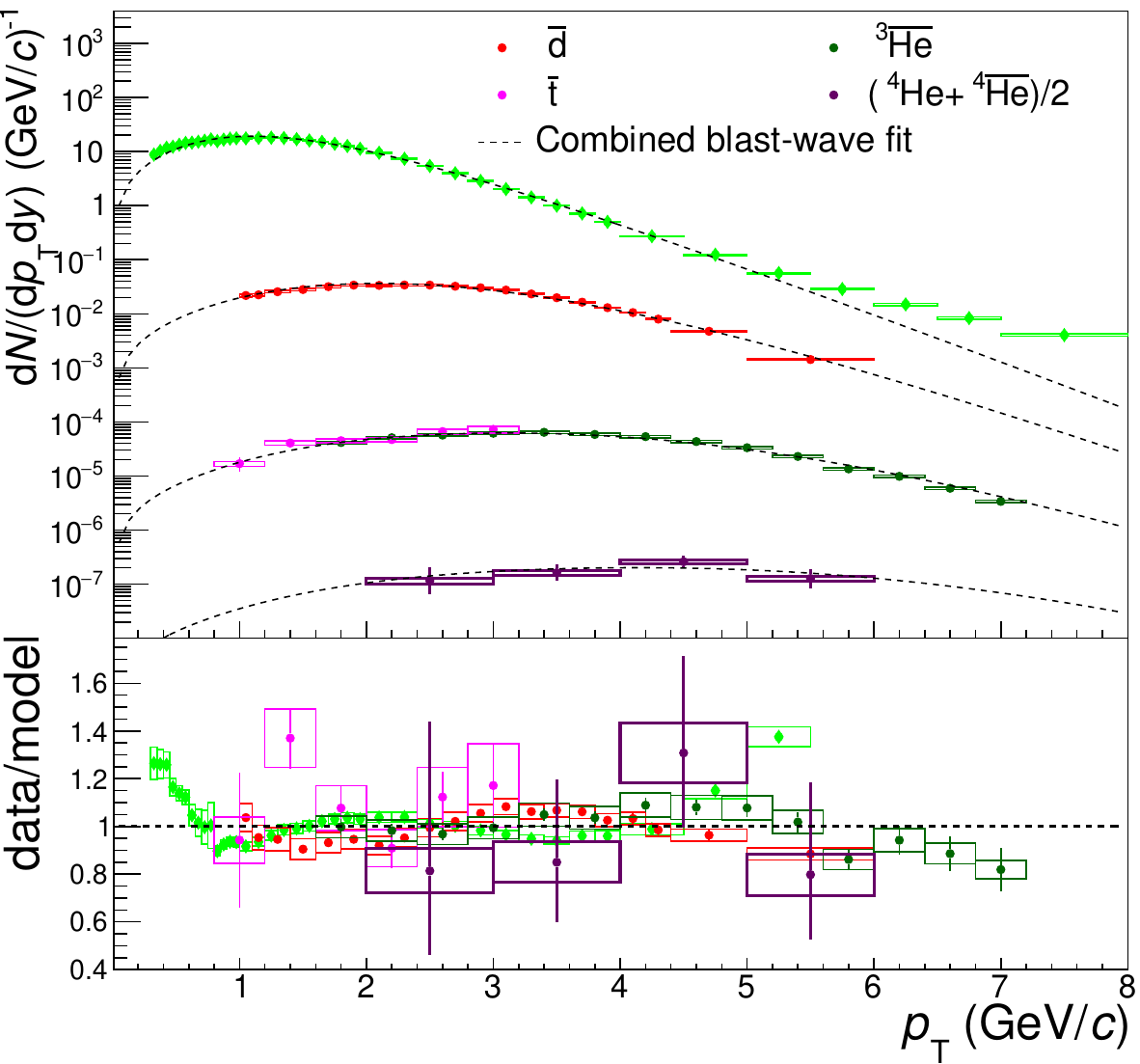}
    \end{center}
    \caption{(Color online) Combined blast-wave fit of all available light flavored hadron $p_{\rm T}$ spectra including nuclei~\cite{ALICE:2019hno,Nuclei_5TeV} (left) and only p, $\overline{\mathrm{d}}$, $\overline{\mathrm{t}}$, ${}^{3}\overline{\mathrm{He}}$ and ${}^{4}$He $p_{\rm T}$ spectra (right) in Pb--Pb collisions at $\sqrt{s_{\rm NN}} = 5.02$ TeV for 0--10\% central events (upper panels). The lower panels show the ratio between each data point and the blast-wave model fit for each species.}
    \label{fig:bw_nucl}
\end{figure}

The combined (anti)alpha \pt spectrum was compared to those of other light-flavored hadrons~\cite{ALICE:2019hno} and nuclei~\cite{Nuclei_5TeV}, measured 
in central (0-10$\%$) Pb--Pb collisions at \snn=\,5.02 TeV, by performing a simultaneous blast-wave fit to all \pt spectra (see Fig.~\ref{fig:bw_nucl}, left). 
The fit range of $\pi$, K, p was restricted in the momentum range in order to minimize biases from resonance decays at low \pt and
from hard processes at high \pt. The fit is performed in the following \pt intervals: 0.5–1 GeV/$c$ for charged pions, 0.2–1.5 GeV/c for charged kaons, and 0.3–3 GeV/c for (anti)protons. These regions are the same as in the previous publications that showed results for global blast-wave fits~\cite{Abelev:2013vea,Adam:2015vda,ALICE:2019hno}. The spectra of antideuterons, antitritons, $^3\overline{\rm{He}}$, and alpha were fitted over the full measured \pt range. 

One should note, blast-wave fits are a simplified approach mimicking the hydrodynamics behind the radial expansion and have certain limitations, e.g.~it is known that the temperature is particularly sensitive to the fit range and the used particle species. In particular, in blast-wave fits using the FastReso package~\cite{Mazeliauskas:2018irt,fastreso} the quality of the fits is rather good using a single temperature of about 150 MeV for chemical and kinetic freeze out and these fits do not show a dependence of the temperature on centrality~\cite{Mazeliauskas:2019ifr}. This is possible in the FastReso approach because the feed-down from resonances is taken into account by the package. 
In addition, there are other approaches utilizing results from LHC that can describe the data in an extended blast-wave model approach with more parameters~\cite{Grigoryan:2021org}. In any case, the standard (Boltzmann-Gibbs) blast-wave fit provides a simple and solid approach to compare the spectra of nuclei and lighter hadrons, which is the goal of the study presented here.

The fit results are shown in the left panel of Fig.~\ref{fig:bw_nucl} and the fit parameters are reported in Table~\ref{tab:bw} (Fit A). The freeze-out parameters, in particular $\langle\beta\rangle$ and $T_{\rm kin}$, are consistent with those obtained in Pb--Pb collisions
at \snn = 2.76 TeV~\cite{Adam:2015vda}. The data-to-model ratios, shown in the bottom panel of Fig.~\ref{fig:bw_nucl} left, indicate that the spectra of nuclei are reasonably well described by the common fit within their uncertainties. This suggests that also relatively heavy compound objects like (anti)alpha nuclei participate in a common flow field.

\begin{table}[!h]
	\centering
	\caption{Parameters obtained from the combined blast-wave fits (Fig.~\ref{fig:bw_nucl}) to the \pt spectra of different combinations of light-flavor hadrons and nuclei measured in central (0--10\%) Pb--Pb collisions at $\sqrt{s_{\rm NN}} = $5.02 TeV. The uncertainty from the fits corresponds to the statistical uncertainty. Systematics, that have been evaluated by changing the fit strategy slightly, are of similar size as the statistical uncertainties. The last column shows the $\chi^2$ value and the corresponding number of degrees of freedom (ndf) for each fit. 
 }
	\begin{tabular}{|l|l|c|c|c|c|c|}
	\hline
	&Fitted particles & $\langle\beta\rangle$ & $\beta_{\rm max}$ & $T_{\rm kin}$ (MeV) & $n$  & $\chi^2 /$ ndf \\
	\hline
  Fit A&$\pi$,K,p,d,t,$^3$He,$^4$He & $0.664 \pm 0.002 $ & $0.873 \pm 0.004 $ & $108 \pm 2$ & $0.63 \pm 0.02$ & 381.1 / 92 \\
	Fit B &p,d,t,${}^{3}$He,${}^{4}$He & $0.670 \pm 0.002 $ & $0.853 \pm 0.004 $ & $132 \pm 4 $ & $0.55 \pm 0.02 $ & 176.5 / 64 \\
 Fit C&d,t,${}^{3}$He,${}^{4}$He & $0.684 \pm 0.003 $ & $0.863 \pm 0.005 $ & $108 \pm 6$ & $0.52 \pm 0.02 $ & 44.5 / 37\\
	Fit D &$\pi$,K,p & $0.664 \pm 0.002 $ & $0.909 \pm 0.003 $ & $ 85 \pm 4$ & $0.74 \pm 0.01 $ & 113.0 / 54 \\
     \hline
		\end{tabular}
	\label{tab:bw}
\end{table}

The coalescence picture assumes that nuclei are formed at a late stage of the collision, i.e.~at or after kinetic freeze-out.
In this case, one may expect that the \pt spectra of protons and (anti)nuclei exhibit a common temperature and velocity field
that characterizes the source at or after the stage of nuclear cluster formation.
To elucidate this further a blast-wave fit was performed, where only protons and (anti)nuclei are included (Fit B). 
The data points are well described by the common fit, as shown in Fig.~\ref{fig:bw_nucl} (right). Actually, the protons are well described over a larger range in Fit B (right panel of Fig.~\ref{fig:bw_nucl}) than in Fit A. The fit parameters indicate a similar velocity field
as in the case when $\pi$ and K are included in the fit (Fit A), but a significantly larger kinetic freeze-out temperature of $T_{\rm kin} = (132 \pm 4)$ MeV.
In the context of final-state coalescence, this finding is unexpected. However, it matches the conjecture of statistical hadronization including formation of (anti)nuclei close to the phase boundary,  without significant rescattering at later stages of the system evolution. 
Possible explanations for such a scenario in terms of pre-hadronic multi-quark states have been proposed in Ref.~\cite{Andronic:2017pug}.

The result is challenged by a fit to only the (anti)nuclei (Fit C) which yields $T_{\rm kin}$ = (108 $\pm$ 6) MeV, which is consistent within the uncertainties with the result of Fit A. This seems to be more in agreement with the expectation of the coalescence model, namely that the protons freeze out earlier as suggested by Fit B, i.e. at a higher temperature, and the nuclei are formed later from these protons and neutrons available for the coalescence process.
A fit to only $\pi$, K, p (Fit D) results in $T_{\rm kin}$ = (85 $\pm$ 4) MeV, indicating that very low apparent kinetic freeze-out temperatures are driven by the lightest particles. It should be noted that lighter particles are more prone to contributions from resonance decays and hard scatterings over a wider \pt range than heavier particles.

From the quality of the fits, i.e. the $\chi^2/$ndf values given in~Table~\ref{tab:bw}, it seems like the separation into nuclei (Fit C) and light-flavored hadrons (Fit D) is best. Nevertheless, the temperature of the latter is lower than Fit C, so the coalescence picture is again questioned from this inconsistency between blast-wave results. Indeed, the temperatures extracted from the fits would imply that the protons used in the coalescence process freeze out later than the nuclei formed from them.\\
Clearly, these findings cannot be used for any strong conclusion, in particular since the blast-wave model is only a simplified hydrodynamical picture that has certain limits as discussed above. 

The rapidity densities d$N$/d$y$ are estimated by integration over the blast-wave function fitted to the limited range of \pt spectra. To this end, a blast-wave fit was performed to the \pt spectra of all particles except (anti)alpha. The resulting fit parameters are used to constrain the shape for (anti)alpha while the normalization is obtained by a fit to the (anti)alpha distributions. This procedure was applied separately for the alpha 
and antialpha \pt distributions as well as to the combined spectrum. The derived rapidity densities are summarized in Table~\ref{tab:dndy}. The statistical
uncertainties are those of the normalization from the fit, while the systematic uncertainties reflect the variation of d$N$/d$y$
if the data points are shifted by their systematic uncertainties. The results for alpha and antialpha are consistent within their uncertainties. Also reported are the SHM results obtained from a fit of all available hadron yields using a grand-canonical ensemble~\cite{Andronic:2010qu,Vovchenko:2019pjl,Vovchenko:2018fiy}.

\begin{table}[!h]
	\centering
	\caption{Rapidity densities of ${}^{4}$He and ${}^{4}\overline{{\rm He}}$ and their average, together with the statistical hadronization model predictions~\cite{Andronic:2010qu,Vovchenko:2019pjl,Vovchenko:2018fiy}. The experimental values are stated with statistical (second value) and systematic uncertainties (third value). }
	\begin{tabular}{|l|c|c|c|}
	\hline
	d$N$/d$y$ (${10}^{-6}$):	& $^4\overline{\mathrm{He}}$  & $^4$He & ($^4\overline{\mathrm{He}} + ^4$He)/2   \\
	\hline
 Experiment & $(1.30\pm 0.28\pm 0.18)$  & $(0.83 \pm 0.22 \pm 0.12)$  &  $(1.00 \pm 0.19 \pm 0.10)$  \\
    SHM ($T_{\mathrm{ch}} = $156 MeV)  &  0.945 & 0.949 & 0.947 \\
     \hline
		\end{tabular}
	\label{tab:dndy}
\end{table}

\begin{figure}[tbh]
    \begin{center}
    \includegraphics[width = 0.7\textwidth]{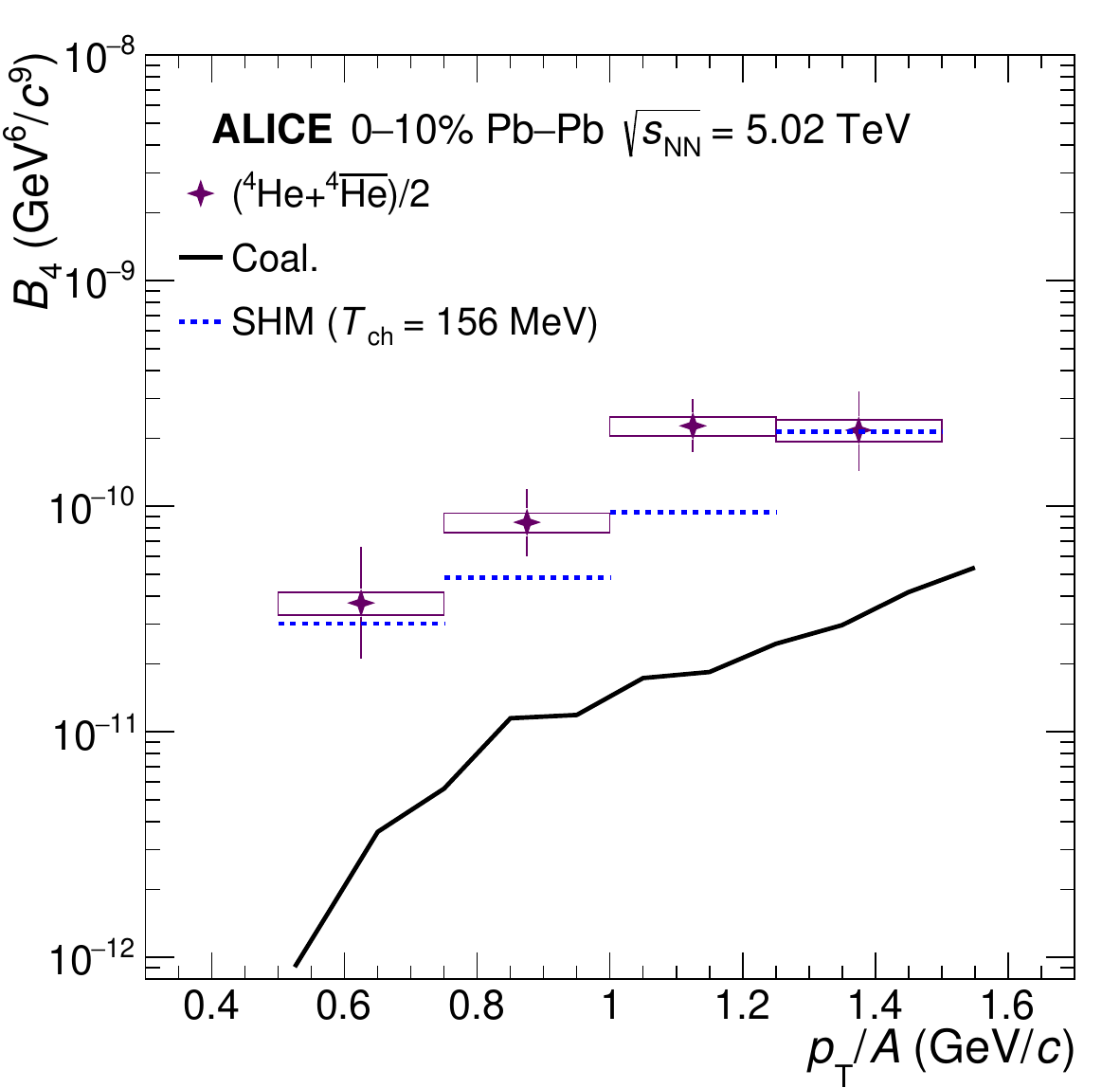}
    \end{center}
    \caption{(Color online) The coalescence parameter $B_4$ as a function of $p_{\rm T}/A$, calculated from the averaged ${}^{4}$He and \antihe spectra and the protons from~\cite{ALICE:2019hno}. Statistical uncertainties are indicated by the vertical lines and the boxes correspond to the systematic uncertainties. The blue dashed line and the full black line indicate the values for the SHM combined with blast-wave \pt shapes and the coalescence predictions from Refs.~\cite{Bellini:2018epz,Bellini:2019zqc}, respectively.}
    \label{fig:b4}
\end{figure}

The presented (anti)alpha transverse-momentum spectra allow for the first time a determination of the coalescence parameter $B_4$ 
at LHC energies. To this end, Eq.~\ref{eq:coalescence} was employed where the proton \pt distributions were taken from Ref.~\cite{ALICE:2019hno} after averaging the measurements in the 0--5\% and 5--10\% centrality intervals.
The $B_4$ values shown in Fig.~\ref{fig:b4} exhibit an increasing trend with \pt/$A$, which is the transverse momentum per nucleon. This trend is similar to earlier measurements in heavy-ion
collisions for lighter nuclei~\cite{Adam:2015vda,Nuclei_5TeV}.
The results in Fig.~\ref{fig:b4} are compared to predictions from coalescence\,\cite{Bellini:2019zqc} and from statistical hadronization models.
For the latter, the (anti)alpha and proton yields (d$N$/d$y$) are calculated for a chemical freeze-out temperature of $T_{\rm{ch}}$ = 156 MeV and the shapes of the transverse-momentum distributions are taken from the blast-wave fit.
While SHM, combined with the spectral shape derived from the blast-wave fit, slightly underpredicts the data, the coalescence prediction is about one order of magnitude below the data in all \pt intervals. However, both models capture the increase of the data well. So intrinsically, the spectral shape seems to be correct in both approaches and the magnitude of the discrepancy between the coalescence curve and the data needs to be understood better.

\begin{figure}[tbh]
    \begin{center}
    \includegraphics[width = 0.8\textwidth]{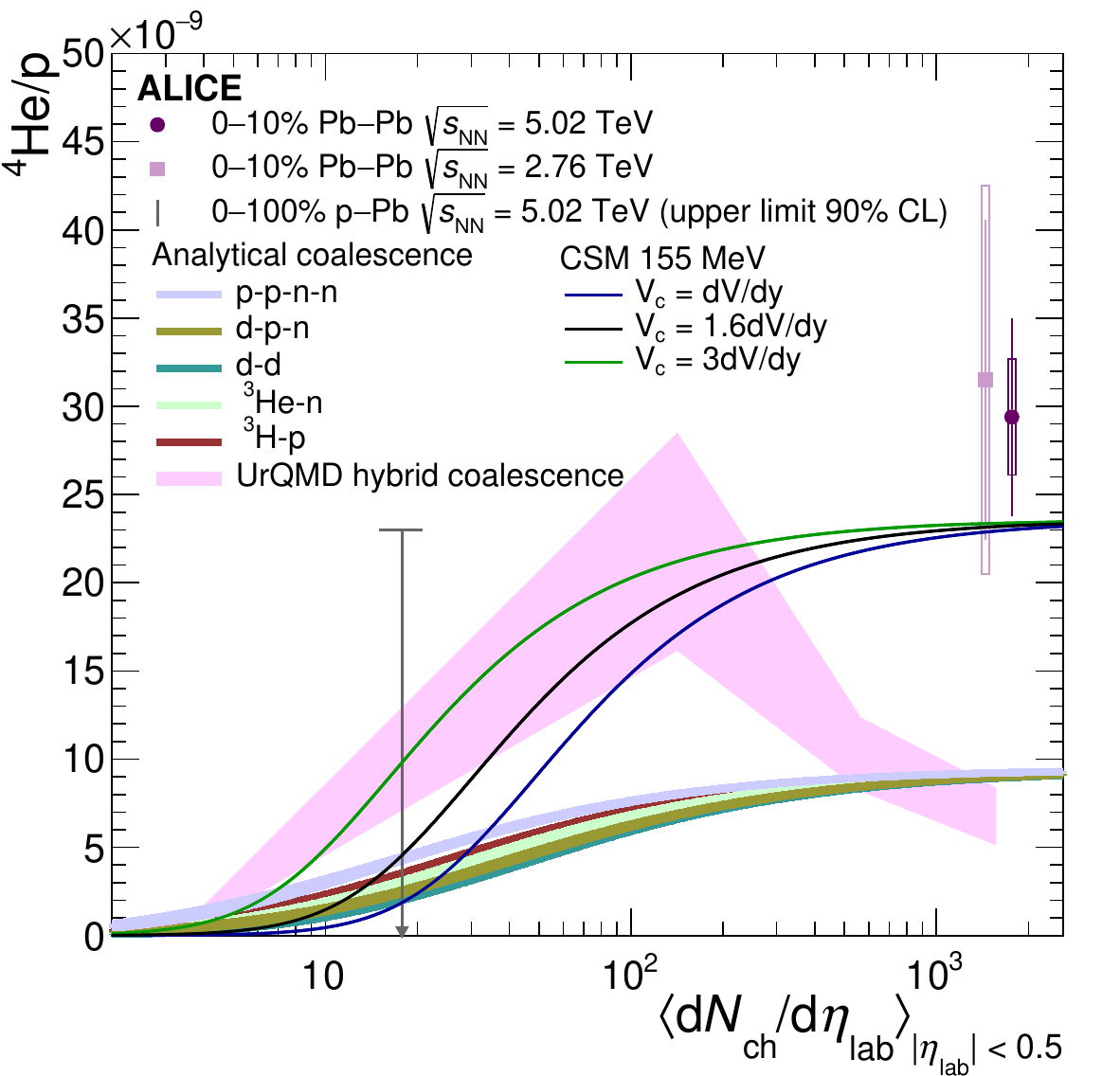}
    \end{center}
    \caption{(Color online) $^4$He/p ratios for the measured data points as a function of charged-particle multiplicity $\langle {\rm d}N_{\rm ch}/{\rm d}\eta \rangle_{|\eta_{\rm lab}|<0.5}$ compared to model predictions. For comparison the result from the 10\% most central Pb--Pb collisions at $\sqrt{s_\mathrm{NN}} = 2.76$ TeV~\cite{Acharya:2017bso} and the upper limit at 90\% CL from p--Pb collisions at $\sqrt{s_\mathrm{NN}} = 5.02$ TeV~\cite{Acharya:2019xmu} is also shown. The thermal model curves are from the CSM~\cite{Vovchenko:2018fiy}. For the coalescence model two different approaches are displayed: analytical and UrQMD hybrid coalescence~\cite{jan,kjs}. The analytical coalescence is shown for five different substructures and the thickness of the bands reflects the uncertainties of the calculation. For the UrQMD model the band is representing the statistical uncertainty of the prediction.}
    \label{fig:ratio}
\end{figure}

The ratio of alpha to proton d$N$/d$y$ in central Pb--Pb collisions at \snn = 5.02 TeV is shown in Fig.~\ref{fig:ratio} as a function of the pseudorapidity density of charged particles produced at midrapidity in the collision, $\langle \mathrm{d}N_{\mathrm{ch}}/\mathrm{d}\eta \rangle_{|\eta_{\mathrm{lab}} |\,<\, 0.5}$. 
In addition, the ratio from the 10\% most central Pb--Pb collisions at \snn = 2.76 TeV~\cite{Acharya:2017bso} and the upper limit in p--Pb collisions at \snn = 5.02 TeV~\cite{Acharya:2019xmu} are depicted.
The new result agrees well with the measurement at lower energy~\cite{Acharya:2017bso}.
Furthermore, predictions from the canonical statistical model (CSM) for $T_{\rm{ch}}$ = 155 MeV and 
three different values of the correlation volume $V_{\rm{C}}$ are displayed~\cite{Vovchenko:2018fiy}. The curves differ at low $\langle \mathrm{d}N_{\mathrm{ch}}/\mathrm{d}\eta \rangle$, corresponding to the multiplicity of charged particles produced in small collision systems, 
but coincide in central Pb--Pb collisions where they are consistent within uncertainties with the measurements. 
Also shown are different calculations from coalescence models. The ``box coalescence" (using a maximal difference in coordinate space and momentum for the coalescing partners) implemented in the UrQMD~\cite{Reichert:2022mek} model, indicated by the magenta band,
shows a non-monotonic behavior that can be explained by absorption processes in the hadronic phase of Pb--Pb collisions ~\cite{jan}. In central Pb--Pb collisions, the UrQMD hybrid model underestimates the data by about a factor of three. The same trend is observed in a CSM approach that includes annihilations~\cite{Vovchenko:2022xil}, which also underestimates the data.
Finally, calculations of an analytical coalescence approach are presented, in which the internal structure of the alpha nucleus is taken into account~\cite{kjs}. 
The assumption of a structureless alpha particle (p-p-n-n) and calculations considering a d-p-n, d-d, ${}^{3}$H-p or ${}^{3}$He-n substructures 
are compared. All analytical coalescence curves coincide for large system sizes where they underestimate the data by about a factor of three. 
This might be connected to the fact that the binding energy of the alpha is not taken into account in the model. Neglecting the binding energy might be working well for the $A = 2$ and $A = 3$ nuclei, but not for the alpha, since it is much tighter bound compared to lighter nuclei ($E_B = 2.2$ MeV for the deuteron, whereas the alpha is bound by $28.3$ MeV). A recent publication determined the mass radius of the alpha to be (1.70 $\pm$ 0.14) fm in a model dependent approach using $\phi(1020)$-photoproduction data, that is close to the measured charge radius which is not expected~\cite{Wang:2023uek}. The mass radius of the alpha used in the coalescence calculations is rather 1.4 fm, estimated from the measured charge radius of (1.6755 $\pm$ 0.0028) fm~\cite{Angeli:2013epw,Bellini:2019zqc}. In fact, there is a more precise measurement of the charge radius using laser spectroscopy of muonic helium ions, that gives (1.67824 $\pm$ 0.00083) fm~\cite{Krauth:2021foz}.  
It should be noted that, the presented data even allows for the sum of contributions from coalescence and statistical hadronization predictions. Since these processes are not mutually exclusive one could actually imagine it as interplay of these two production mechanisms.

\section{Conclusion}
\label{sec:con}

New results on (anti)alpha production in central Pb--Pb collisions at \snn = 5.02 TeV were presented, including the first
differential measurement of the antialpha transverse-momentum distribution. 
Predictions from statistical hadronization models are compatible with the measured coalescence parameters $B_4$ and the
(anti)alpha-to-proton yield ratio. In contrast, different implementations of the coalescence model underestimate the data significantly. These findings for the production of (anti)alpha are different from the results for $A = 3$ nuclei~\cite{Nuclei_5TeV}, where both classes of models differ only by about 30$\%$ and the data tend to lie in between. Improvements for the models, e.g.\ incorporating the binding energy of the alpha, are needed to get a better understanding of its production.    
A blast-wave analysis of the \pt distributions together with other hadrons and light nuclei
from central Pb--Pb collisions suggests that also relatively heavy compound objects like (anti)alpha nuclei participate in a common
flow field. However, the constraint of the (anti)alpha on this is limited by the current statistics.
On the other hand, a blast-wave fit including only protons and light nuclei up to \he results in a kinetic freeze-out
temperature that is rather close to the chemical freeze-out temperature obtained from statistical hadronization models. Note that one should be careful with any strong conclusion from the blast-wave fit, since it has certain limitations, e.g.\ being sensitive on the fit regions and the treatment of feed-down from resonances.  
Nevertheless, this result does not agree with na\"{\i}ve expectations based on the coalescence picture, but is in line with a scenario where the yields of light nuclei in central Pb--Pb collisions are dominated by thermal production close to the QCD phase boundary.  
It should be noted that thermal production and coalescence are not mutually exclusive processes and that the data presented here are
even compatible with the sum of contributions from coalescence and statistical hadronization, suggesting a possible interplay of these two production mechanisms.

The recent upgrades of the ALICE detector will enable the collection of substantially larger data samples during LHC Runs 3 and 4. 
This will allow for more differential measurements of (anti)alpha production, enabling in particular a systematic study of its dependence on multiplicity and collision system size. The large sensitivity of the (anti)alpha yield to the different production scenarios may help to shed
light on the interplay between coalescence and thermal production and a possible transition between them
at intermediate system sizes.


\newenvironment{acknowledgement}{\relax}{\relax}
\begin{acknowledgement}
\section*{Acknowledgements}
We thank J. Steinheimer and K.-J. Sun for useful discussions and for providing their predictions.

The ALICE Collaboration would like to thank all its engineers and technicians for their invaluable contributions to the construction of the experiment and the CERN accelerator teams for the outstanding performance of the LHC complex.
The ALICE Collaboration gratefully acknowledges the resources and support provided by all Grid centres and the Worldwide LHC Computing Grid (WLCG) collaboration.
The ALICE Collaboration acknowledges the following funding agencies for their support in building and running the ALICE detector:
A. I. Alikhanyan National Science Laboratory (Yerevan Physics Institute) Foundation (ANSL), State Committee of Science and World Federation of Scientists (WFS), Armenia;
Austrian Academy of Sciences, Austrian Science Fund (FWF): [M 2467-N36] and Nationalstiftung f\"{u}r Forschung, Technologie und Entwicklung, Austria;
Ministry of Communications and High Technologies, National Nuclear Research Center, Azerbaijan;
Conselho Nacional de Desenvolvimento Cient\'{\i}fico e Tecnol\'{o}gico (CNPq), Financiadora de Estudos e Projetos (Finep), Funda\c{c}\~{a}o de Amparo \`{a} Pesquisa do Estado de S\~{a}o Paulo (FAPESP) and Universidade Federal do Rio Grande do Sul (UFRGS), Brazil;
Bulgarian Ministry of Education and Science, within the National Roadmap for Research Infrastructures 2020-2027 (object CERN), Bulgaria;
Ministry of Education of China (MOEC) , Ministry of Science \& Technology of China (MSTC) and National Natural Science Foundation of China (NSFC), China;
Ministry of Science and Education and Croatian Science Foundation, Croatia;
Centro de Aplicaciones Tecnol\'{o}gicas y Desarrollo Nuclear (CEADEN), Cubaenerg\'{\i}a, Cuba;
Ministry of Education, Youth and Sports of the Czech Republic, Czech Republic;
The Danish Council for Independent Research | Natural Sciences, the VILLUM FONDEN and Danish National Research Foundation (DNRF), Denmark;
Helsinki Institute of Physics (HIP), Finland;
Commissariat \`{a} l'Energie Atomique (CEA) and Institut National de Physique Nucl\'{e}aire et de Physique des Particules (IN2P3) and Centre National de la Recherche Scientifique (CNRS), France;
Bundesministerium f\"{u}r Bildung und Forschung (BMBF) and GSI Helmholtzzentrum f\"{u}r Schwerionenforschung GmbH, Germany;
General Secretariat for Research and Technology, Ministry of Education, Research and Religions, Greece;
National Research, Development and Innovation Office, Hungary;
Department of Atomic Energy Government of India (DAE), Department of Science and Technology, Government of India (DST), University Grants Commission, Government of India (UGC) and Council of Scientific and Industrial Research (CSIR), India;
National Research and Innovation Agency - BRIN, Indonesia;
Istituto Nazionale di Fisica Nucleare (INFN), Italy;
Japanese Ministry of Education, Culture, Sports, Science and Technology (MEXT) and Japan Society for the Promotion of Science (JSPS) KAKENHI, Japan;
Consejo Nacional de Ciencia (CONACYT) y Tecnolog\'{i}a, through Fondo de Cooperaci\'{o}n Internacional en Ciencia y Tecnolog\'{i}a (FONCICYT) and Direcci\'{o}n General de Asuntos del Personal Academico (DGAPA), Mexico;
Nederlandse Organisatie voor Wetenschappelijk Onderzoek (NWO), Netherlands;
The Research Council of Norway, Norway;
Commission on Science and Technology for Sustainable Development in the South (COMSATS), Pakistan;
Pontificia Universidad Cat\'{o}lica del Per\'{u}, Peru;
Ministry of Education and Science, National Science Centre and WUT ID-UB, Poland;
Korea Institute of Science and Technology Information and National Research Foundation of Korea (NRF), Republic of Korea;
Ministry of Education and Scientific Research, Institute of Atomic Physics, Ministry of Research and Innovation and Institute of Atomic Physics and Universitatea Nationala de Stiinta si Tehnologie Politehnica Bucuresti, Romania;
Ministry of Education, Science, Research and Sport of the Slovak Republic, Slovakia;
National Research Foundation of South Africa, South Africa;
Swedish Research Council (VR) and Knut \& Alice Wallenberg Foundation (KAW), Sweden;
European Organization for Nuclear Research, Switzerland;
Suranaree University of Technology (SUT), National Science and Technology Development Agency (NSTDA) and National Science, Research and Innovation Fund (NSRF via PMU-B B05F650021), Thailand;
Turkish Energy, Nuclear and Mineral Research Agency (TENMAK), Turkey;
National Academy of  Sciences of Ukraine, Ukraine;
Science and Technology Facilities Council (STFC), United Kingdom;
National Science Foundation of the United States of America (NSF) and United States Department of Energy, Office of Nuclear Physics (DOE NP), United States of America.
In addition, individual groups or members have received support from:
Czech Science Foundation (grant no. 23-07499S), Czech Republic;
European Research Council, Strong 2020 - Horizon 2020 (grant nos. 950692, 824093), European Union;
ICSC - Centro Nazionale di Ricerca in High Performance Computing, Big Data and Quantum Computing, European Union - NextGenerationEU;
Academy of Finland (Center of Excellence in Quark Matter) (grant nos. 346327, 346328), Finland.

\end{acknowledgement}

\bibliographystyle{utphys}   
\bibliography{bibliography}

\newpage
\appendix

%
%

\section{The ALICE Collaboration}
\label{app:collab}
\begin{flushleft} 
\small

S.~Acharya\,\orcidlink{0000-0002-9213-5329}\,$^{\rm 128}$, 
D.~Adamov\'{a}\,\orcidlink{0000-0002-0504-7428}\,$^{\rm 87}$, 
G.~Aglieri Rinella\,\orcidlink{0000-0002-9611-3696}\,$^{\rm 33}$, 
M.~Agnello\,\orcidlink{0000-0002-0760-5075}\,$^{\rm 30}$, 
N.~Agrawal\,\orcidlink{0000-0003-0348-9836}\,$^{\rm 52}$, 
Z.~Ahammed\,\orcidlink{0000-0001-5241-7412}\,$^{\rm 136}$, 
S.~Ahmad\,\orcidlink{0000-0003-0497-5705}\,$^{\rm 16}$, 
S.U.~Ahn\,\orcidlink{0000-0001-8847-489X}\,$^{\rm 72}$, 
I.~Ahuja\,\orcidlink{0000-0002-4417-1392}\,$^{\rm 38}$, 
A.~Akindinov\,\orcidlink{0000-0002-7388-3022}\,$^{\rm 142}$
% ...

\section*{Affiliation Notes}

$^{\rm I}$ Deceased\\
$^{\rm II}$ Also at: Max-Planck-Institut fur Physik, Munich, Germany\\
$^{\rm III}$ Also at: Italian National Agency for New Technologies, Energy and Sustainable Economic Development (ENEA), Bologna, Italy\\
$^{\rm IV}$ Also at: Dipartimento DET del Politecnico di Torino, Turin, Italy\\
$^{\rm V}$ Also at: Yildiz Technical University, Istanbul, T\"{u}rkiye\\
$^{\rm VI}$ Also at: Department of Applied Physics, Aligarh Muslim University, Aligarh, India\\
$^{\rm VII}$ Also at: Institute of Theoretical Physics, University of Wroclaw, Poland\\
$^{\rm VIII}$ Also at: An institution covered by a cooperation agreement with CERN\\

\section*{Collaboration Institutes}

$^{1}$ A.I. Alikhanyan National Science Laboratory (Yerevan Physics Institute) Foundation, Yerevan, Armenia\\
$^{2}$ AGH University of Krakow, Cracow, Poland\\
$^{3}$ Bogolyubov Institute for Theoretical Physics, National Academy of Sciences of Ukraine, Kiev, Ukraine\\
$^{4}$ Bose Institute, Department of Physics  and Centre for Astroparticle Physics and Space Science (CAPSS), Kolkata, India\\
$^{5}$ California Polytechnic State University, San Luis Obispo, California, United States\\
$^{6}$ Central China Normal University, Wuhan, China\\
$^{7}$ Centro de Aplicaciones Tecnol\'{o}gicas y Desarrollo Nuclear (CEADEN), Havana, Cuba\\
$^{8}$ Centro de Investigaci\'{o}n y de Estudios Avanzados (CINVESTAV), Mexico City and M\'{e}rida, Mexico\\
$^{9}$ Chicago State University, Chicago, Illinois, United States\\
$^{10}$ China Institute of Atomic Energy, Beijing, China\\
$^{11}$ China University of Geosciences, Wuhan, China\\
$^{12}$ Chungbuk National University, Cheongju, Republic of Korea\\
$^{13}$ Comenius University Bratislava, Faculty of Mathematics, Physics and Informatics, Bratislava, Slovak Republic\\
$^{14}$ COMSATS University Islamabad, Islamabad, Pakistan\\
$^{15}$ Creighton University, Omaha, Nebraska, United States\\
$^{16}$ Department of Physics, Aligarh Muslim University, Aligarh, India\\
$^{17}$ Department of Physics, Pusan National University, Pusan, Republic of Korea\\
$^{18}$ Department of Physics, Sejong University, Seoul, Republic of Korea\\
$^{19}$ Department of Physics, University of California, Berkeley, California, United States\\
$^{20}$ Department of Physics, University of Oslo, Oslo, Norway\\
$^{21}$ Department of Physics and Technology, University of Bergen, Bergen, Norway\\
$^{22}$ Dipartimento di Fisica, Universit\`{a} di Pavia, Pavia, Italy\\
$^{23}$ Dipartimento di Fisica dell'Universit\`{a} and Sezione INFN, Cagliari, Italy\\
$^{24}$ Dipartimento di Fisica dell'Universit\`{a} and Sezione INFN, Trieste, Italy\\
$^{25}$ Dipartimento di Fisica dell'Universit\`{a} and Sezione INFN, Turin, Italy\\
$^{26}$ Dipartimento di Fisica e Astronomia dell'Universit\`{a} and Sezione INFN, Bologna, Italy\\
$^{27}$ Dipartimento di Fisica e Astronomia dell'Universit\`{a} and Sezione INFN, Catania, Italy\\
$^{28}$ Dipartimento di Fisica e Astronomia dell'Universit\`{a} and Sezione INFN, Padova, Italy\\
$^{29}$ Dipartimento di Fisica `E.R.~Caianiello' dell'Universit\`{a} and Gruppo Collegato INFN, Salerno, Italy\\
$^{30}$ Dipartimento DISAT del Politecnico and Sezione INFN, Turin, Italy\\
$^{31}$ Dipartimento di Scienze MIFT, Universit\`{a} di Messina, Messina, Italy\\
$^{32}$ Dipartimento Interateneo di Fisica `M.~Merlin' and Sezione INFN, Bari, Italy\\
$^{33}$ European Organization for Nuclear Research (CERN), Geneva, Switzerland\\
$^{34}$ Faculty of Electrical Engineering, Mechanical Engineering and Naval Architecture, University of Split, Split, Croatia\\
$^{35}$ Faculty of Engineering and Science, Western Norway University of Applied Sciences, Bergen, Norway\\
$^{36}$ Faculty of Nuclear Sciences and Physical Engineering, Czech Technical University in Prague, Prague, Czech Republic\\
$^{37}$ Faculty of Physics, Sofia University, Sofia, Bulgaria\\
$^{38}$ Faculty of Science, P.J.~\v{S}af\'{a}rik University, Ko\v{s}ice, Slovak Republic\\
$^{39}$ Frankfurt Institute for Advanced Studies, Johann Wolfgang Goethe-Universit\"{a}t Frankfurt, Frankfurt, Germany\\
$^{40}$ Fudan University, Shanghai, China\\
$^{41}$ Gangneung-Wonju National University, Gangneung, Republic of Korea\\
$^{42}$ Gauhati University, Department of Physics, Guwahati, India\\
$^{43}$ Helmholtz-Institut f\"{u}r Strahlen- und Kernphysik, Rheinische Friedrich-Wilhelms-Universit\"{a}t Bonn, Bonn, Germany\\
$^{44}$ Helsinki Institute of Physics (HIP), Helsinki, Finland\\
$^{45}$ High Energy Physics Group,  Universidad Aut\'{o}noma de Puebla, Puebla, Mexico\\
$^{46}$ Horia Hulubei National Institute of Physics and Nuclear Engineering, Bucharest, Romania\\
$^{47}$ HUN-REN Wigner Research Centre for Physics, Budapest, Hungary\\
$^{48}$ Indian Institute of Technology Bombay (IIT), Mumbai, India\\
$^{49}$ Indian Institute of Technology Indore, Indore, India\\
$^{50}$ INFN, Laboratori Nazionali di Frascati, Frascati, Italy\\
$^{51}$ INFN, Sezione di Bari, Bari, Italy\\
$^{52}$ INFN, Sezione di Bologna, Bologna, Italy\\
$^{53}$ INFN, Sezione di Cagliari, Cagliari, Italy\\
$^{54}$ INFN, Sezione di Catania, Catania, Italy\\
$^{55}$ INFN, Sezione di Padova, Padova, Italy\\
$^{56}$ INFN, Sezione di Pavia, Pavia, Italy\\
$^{57}$ INFN, Sezione di Torino, Turin, Italy\\
$^{58}$ INFN, Sezione di Trieste, Trieste, Italy\\
$^{59}$ Inha University, Incheon, Republic of Korea\\
$^{60}$ Institute for Gravitational and Subatomic Physics (GRASP), Utrecht University/Nikhef, Utrecht, Netherlands\\
$^{61}$ Institute of Experimental Physics, Slovak Academy of Sciences, Ko\v{s}ice, Slovak Republic\\
$^{62}$ Institute of Physics, Homi Bhabha National Institute, Bhubaneswar, India\\
$^{63}$ Institute of Physics of the Czech Academy of Sciences, Prague, Czech Republic\\
$^{64}$ Institute of Space Science (ISS), Bucharest, Romania\\
$^{65}$ Institut f\"{u}r Kernphysik, Johann Wolfgang Goethe-Universit\"{a}t Frankfurt, Frankfurt, Germany\\
$^{66}$ Instituto de Ciencias Nucleares, Universidad Nacional Aut\'{o}noma de M\'{e}xico, Mexico City, Mexico\\
$^{67}$ Instituto de F\'{i}sica, Universidade Federal do Rio Grande do Sul (UFRGS), Porto Alegre, Brazil\\
$^{68}$ Instituto de F\'{\i}sica, Universidad Nacional Aut\'{o}noma de M\'{e}xico, Mexico City, Mexico\\
$^{69}$ iThemba LABS, National Research Foundation, Somerset West, South Africa\\
$^{70}$ Jeonbuk National University, Jeonju, Republic of Korea\\
$^{71}$ Johann-Wolfgang-Goethe Universit\"{a}t Frankfurt Institut f\"{u}r Informatik, Fachbereich Informatik und Mathematik, Frankfurt, Germany\\
$^{72}$ Korea Institute of Science and Technology Information, Daejeon, Republic of Korea\\
$^{73}$ KTO Karatay University, Konya, Turkey\\
$^{74}$ Laboratoire de Physique Subatomique et de Cosmologie, Universit\'{e} Grenoble-Alpes, CNRS-IN2P3, Grenoble, France\\
$^{75}$ Lawrence Berkeley National Laboratory, Berkeley, California, United States\\
$^{76}$ Lund University Department of Physics, Division of Particle Physics, Lund, Sweden\\
$^{77}$ Nagasaki Institute of Applied Science, Nagasaki, Japan\\
$^{78}$ Nara Women{'}s University (NWU), Nara, Japan\\
$^{79}$ National and Kapodistrian University of Athens, School of Science, Department of Physics , Athens, Greece\\
$^{80}$ National Centre for Nuclear Research, Warsaw, Poland\\
$^{81}$ National Institute of Science Education and Research, Homi Bhabha National Institute, Jatni, India\\
$^{82}$ National Nuclear Research Center, Baku, Azerbaijan\\
$^{83}$ National Research and Innovation Agency - BRIN, Jakarta, Indonesia\\
$^{84}$ Niels Bohr Institute, University of Copenhagen, Copenhagen, Denmark\\
$^{85}$ Nikhef, National institute for subatomic physics, Amsterdam, Netherlands\\
$^{86}$ Nuclear Physics Group, STFC Daresbury Laboratory, Daresbury, United Kingdom\\
$^{87}$ Nuclear Physics Institute of the Czech Academy of Sciences, Husinec-\v{R}e\v{z}, Czech Republic\\
$^{88}$ Oak Ridge National Laboratory, Oak Ridge, Tennessee, United States\\
$^{89}$ Ohio State University, Columbus, Ohio, United States\\
$^{90}$ Physics department, Faculty of science, University of Zagreb, Zagreb, Croatia\\
$^{91}$ Physics Department, Panjab University, Chandigarh, India\\
$^{92}$ Physics Department, University of Jammu, Jammu, India\\
$^{93}$ Physics Program and International Institute for Sustainability with Knotted Chiral Meta Matter (SKCM2), Hiroshima University, Hiroshima, Japan\\
$^{94}$ Physikalisches Institut, Eberhard-Karls-Universit\"{a}t T\"{u}bingen, T\"{u}bingen, Germany\\
$^{95}$ Physikalisches Institut, Ruprecht-Karls-Universit\"{a}t Heidelberg, Heidelberg, Germany\\
$^{96}$ Physik Department, Technische Universit\"{a}t M\"{u}nchen, Munich, Germany\\
$^{97}$ Politecnico di Bari and Sezione INFN, Bari, Italy\\
$^{98}$ Research Division and ExtreMe Matter Institute EMMI, GSI Helmholtzzentrum f\"ur Schwerionenforschung GmbH, Darmstadt, Germany\\
$^{99}$ Saga University, Saga, Japan\\
$^{100}$ Saha Institute of Nuclear Physics, Homi Bhabha National Institute, Kolkata, India\\
$^{101}$ School of Physics and Astronomy, University of Birmingham, Birmingham, United Kingdom\\
$^{102}$ Secci\'{o}n F\'{\i}sica, Departamento de Ciencias, Pontificia Universidad Cat\'{o}lica del Per\'{u}, Lima, Peru\\
$^{103}$ Stefan Meyer Institut f\"{u}r Subatomare Physik (SMI), Vienna, Austria\\
$^{104}$ SUBATECH, IMT Atlantique, Nantes Universit\'{e}, CNRS-IN2P3, Nantes, France\\
$^{105}$ Sungkyunkwan University, Suwon City, Republic of Korea\\
$^{106}$ Suranaree University of Technology, Nakhon Ratchasima, Thailand\\
$^{107}$ Technical University of Ko\v{s}ice, Ko\v{s}ice, Slovak Republic\\
$^{108}$ The Henryk Niewodniczanski Institute of Nuclear Physics, Polish Academy of Sciences, Cracow, Poland\\
$^{109}$ The University of Texas at Austin, Austin, Texas, United States\\
$^{110}$ Universidad Aut\'{o}noma de Sinaloa, Culiac\'{a}n, Mexico\\
$^{111}$ Universidade de S\~{a}o Paulo (USP), S\~{a}o Paulo, Brazil\\
$^{112}$ Universidade Estadual de Campinas (UNICAMP), Campinas, Brazil\\
$^{113}$ Universidade Federal do ABC, Santo Andre, Brazil\\
$^{114}$ Universitatea Nationala de Stiinta si Tehnologie Politehnica Bucuresti, Bucharest, Romania\\
$^{115}$ University of Cape Town, Cape Town, South Africa\\
$^{116}$ University of Derby, Derby, United Kingdom\\
$^{117}$ University of Houston, Houston, Texas, United States\\
$^{118}$ University of Jyv\"{a}skyl\"{a}, Jyv\"{a}skyl\"{a}, Finland\\
$^{119}$ University of Kansas, Lawrence, Kansas, United States\\
$^{120}$ University of Liverpool, Liverpool, United Kingdom\\
$^{121}$ University of Science and Technology of China, Hefei, China\\
$^{122}$ University of South-Eastern Norway, Kongsberg, Norway\\
$^{123}$ University of Tennessee, Knoxville, Tennessee, United States\\
$^{124}$ University of the Witwatersrand, Johannesburg, South Africa\\
$^{125}$ University of Tokyo, Tokyo, Japan\\
$^{126}$ University of Tsukuba, Tsukuba, Japan\\
$^{127}$ Universit\"{a}t M\"{u}nster, Institut f\"{u}r Kernphysik, M\"{u}nster, Germany\\
$^{128}$ Universit\'{e} Clermont Auvergne, CNRS/IN2P3, LPC, Clermont-Ferrand, France\\
$^{129}$ Universit\'{e} de Lyon, CNRS/IN2P3, Institut de Physique des 2 Infinis de Lyon, Lyon, France\\
$^{130}$ Universit\'{e} de Strasbourg, CNRS, IPHC UMR 7178, F-67000 Strasbourg, France, Strasbourg, France\\
$^{131}$ Universit\'{e} Paris-Saclay, Centre d'Etudes de Saclay (CEA), IRFU, D\'{e}partment de Physique Nucl\'{e}aire (DPhN), Saclay, France\\
$^{132}$ Universit\'{e}  Paris-Saclay, CNRS/IN2P3, IJCLab, Orsay, France\\
$^{133}$ Universit\`{a} degli Studi di Foggia, Foggia, Italy\\
$^{134}$ Universit\`{a} del Piemonte Orientale, Vercelli, Italy\\
$^{135}$ Universit\`{a} di Brescia, Brescia, Italy\\
$^{136}$ Variable Energy Cyclotron Centre, Homi Bhabha National Institute, Kolkata, India\\
$^{137}$ Warsaw University of Technology, Warsaw, Poland\\
$^{138}$ Wayne State University, Detroit, Michigan, United States\\
$^{139}$ Yale University, New Haven, Connecticut, United States\\
$^{140}$ Yonsei University, Seoul, Republic of Korea\\
$^{141}$  Zentrum  f\"{u}r Technologie und Transfer (ZTT), Worms, Germany\\
$^{142}$ Affiliated with an institute covered by a cooperation agreement with CERN\\
$^{143}$ Affiliated with an international laboratory covered by a cooperation agreement with CERN.\\

\end{flushleft} 

\end{document}